\documentclass[12pt,a4paper]{article}
\usepackage{color,graphics,amsmath,epsfig,rotating}

\begin{document}

\setlength{\unitlength}{1mm}


\def\ma{m_A}
\def\mhf{m_{1/2}}
\def\m0{m_0}
\def\ra{\rightarrow}
\def\neuto {\tilde\chi_1^0}
\def\mneuto{m_{\tilde{\chi}_1^0}}
\def\stauo{\tilde\tau_1}
\def\staur{\tilde{\tau_R}}
\def\ser{\tilde{e_R}}
\def\smur{\tilde{\mu}_R}
\def\mt{m_t}
\def\mbmb{m_b(m_b)}
\def\mslr{m_{\tilde l_R}}

\def\nn              {\notag}
\def\bce             {\begin{center}}
\def\ece             {\end{center}}

\def\mbf             {\boldmath}

\def\ti              {\tilde}

\def\a               {\alpha}
\def\b               {\beta}
\def\d               {\delta}
\def\D               {\Delta}
\def\g               {\gamma}
\def\G               {\Gamma}
\def\l               {\lambda}
\def\t               {\theta}
\def\s               {\sigma}
\def\S               {\Sigma}
\def\x               {\chi}

\def\sq              {\ti q}
\def\sqL             {\ti q_L^{}}
\def\sqR             {\ti q_R^{}}
\def\sf              {\ti f}

\def\st              {\ti t}
\def\sb              {\ti b}
\def\stau            {\ti \tau}
\def\snu             {\ti \nu}

\def\sqbar  {\Bar{\Tilde q}^{}}
\def\stbar  {{\Bar{\Tilde t}}}
\def\sbbar  {{\Bar{\Tilde b}}}

\def\Pp  {{\cal P}_{\!+}}
\def\Pm  {{\cal P}_{\!-}}

\def\ch              {\ti \x^\pm}
\def\chp             {\ti \x^+}
\def\chm             {\ti \x^-}
\def\nt              {\ti \x^0}

\newcommand{\msq}[1]   {m_{\ti q_{#1}}}
\newcommand{\msf}[1]   {m_{\ti f_{#1}}}
\newcommand{\mst}[1]   {m_{\ti t_{#1}}}
\newcommand{\msb}[1]   {m_{\ti b_{#1}}}
\newcommand{\mstau}[1] {m_{\ti\tau_{#1}}}
\newcommand{\mch}[1]   {m_{\ti \x^\pm_{#1}}}
\newcommand{\mnt}[1]   {m_{\ti \x^0_{#1}}}
\newcommand{\mq}{\mbox{$m_{\tilde{q}}$}}
\newcommand{\mhp}      {m_{H^+}}
\newcommand{\msg}      {m_{\ti g}}

\def\sg              {{\ti g}}
\def\msg             {m_{\sg}}

\def\tW              {\t_{\scriptscriptstyle W}}
\def\tsq             {\t_{\sq}}
\def\tst             {\t_{\st}}
\def\tsb             {\t_{\sb}}
\def\tstau           {\t_{\stau}}
\def\tsf             {\t_{\sf}}
\def\sth             {\sin\t}
\def\cth             {\cos\t}
\def\cst             {\cos\t_{\st}}
\def\csb             {\cos\t_{\sb}}

\def\onehf           {{\textstyle \frac{1}{2}}}
\def\oneth           {{\textstyle \frac{1}{3}}}
\def\twoth           {{\textstyle \frac{2}{3}}}

\def\rzw             {\sqrt{2}}

\def\BR              {{\rm BR}}
\def\mev             {{\rm MeV}}
\def\gev             {{\rm GeV}}
\def\tev             {{\rm TeV}}
\def\fb              {{\rm fb}}
\def\fbi             {{\rm fb}^{-1}}

\def\over            {\overline}
\def\MSbar           {{\overline{\rm MS}}}
\def\DRbar           {{\overline{\rm DR}}}
\def\DR              {{\rm\overline{DR}}}
\def\MS              {{\rm\overline{MS}}}

\def\isajet          {{\tt ISAJET\,7.69}}
\def\softsusy        {{\tt SOFTSUSY\,1.9}}
\def\spheno          {{\tt SPHENO\,2.2.2}}
\def\suspect         {{\tt SUSPECT\,2.3}}
\def\micro           {{\tt micrOMEGAs\,1.3}}
\def\micronew        {{\tt micrOMEGAs\,2.0}}
\def\micromegas       {{\tt micrOMEGAs}}
\def\darksusy        {{\tt DarkSUSY}}
\def\calchep         {{\tt CalcHEP}}
\def\lanhep         {{\tt LanHEP}}
\def\isatools        {{\tt IsaTOOLS}}
\def\nmhdecay        {{\tt NMHDecay}}

\def\isajetnn        {{\tt ISAJET}}
\def\softsusynn      {{\tt SOFTSUSY}}
\def\sphenonn        {{\tt SPHENO}}
\def\suspectnn       {{\tt SUSPECT}}
\def\cygwin     {{\tt Cygwin}}

\def\dMb   {\Delta m_b}

\newcommand{\beqn}{\begin{eqnarray}}
\newcommand\eeqn{\end{eqnarray}}

\def\bsgamma{b\to s\gamma}
\def\bino{\tilde{B}}
\def\wino{\tilde{W}}
\def\higgsino{\tilde{H}}
\def\singlino{\tilde{S}}

\newcommand{\eq}[1]  {\mbox{(\ref{eq:#1})}}
\newcommand{\fig}[1] {Fig.~\ref{fig:#1}}
\newcommand{\Fig}[1] {Figure~\ref{fig:#1}}
\newcommand{\tab}[1] {Table~\ref{tab:#1}}
\newcommand{\Tab}[1] {Table~\ref{tab:#1}}


\bce
{

{\Large\bf  \micronew: a  program to calculate the relic density
of dark matter in a generic model
.} \\[10mm]

{\large   G.~B\'elanger$^1$, F.~Boudjema$^1$,  A.~Pukhov$^2$, A.~Semenov$^3$}\\[4mm]

{\it 1) Laboratoire de Physique Theorique LAPTH, F-74941 Annecy-le-Vieux, France\\

     2) Skobeltsyn Inst. of Nuclear Physics, Moscow State Univ., Moscow 119992,
Russia\\
     3) Joint Institute for Nuclear Research (JINR), 141980, Dubna, Russia  }\\[4mm]

\today
}
\ece

\begin{abstract}
\micronew~ is a code which calculates the  relic density of a
stable massive particle in an arbitrary model.  The underlying
assumption is that there is a conservation law like R-parity in
supersymmetry which guarantees the stability of the lightest odd
particle. The new physics model must be incorporated in the
notation of CalcHEP, a package for the automatic generation of
squared matrix elements. Once this is done, all annihilation and
coannihilation channels are included automatically in any model.
Cross-sections at $v=0$, relevant for indirect detection of dark
matter, are also computed automatically. The package includes
three sample models: the minimal supersymmetric standard model
(MSSM), the MSSM with complex phases and the NMSSM. Extension to
other models, including non supersymmetric models, is described.
\end{abstract}

\section{Introduction}

 Precision cosmological measurements have recently provided
strong evidence that the universe is dominated by dark energy and
also contains a large dark matter component ~\cite{Bennett:2003bz,
Spergel:2003cb,Tegmark:2003ud}. Furthermore, the amount of dark
matter today, the relic density, has been measured with good
precision by WMAP, $0.094<\Omega h^2<0.128$ \cite{Bennett:2003bz,
Spergel:2003cb} . While one can show, based on general arguments
\cite{Jungman:1995df}, that a reasonable value for the relic
density of cold dark matter, can be obtained in any model with a
stable particle that is weakly interacting, the precision reached
allows to actually test in details the implication of the new
physics models that propose a dark matter candidate, thus putting
strong constraints on new models. It is particulary interesting
that models for new physics whose prime goal is to solve the
hierarchy problem, for example supersymmetry, can in many cases
also provide a viable candidate for dark matter. In the minimal
supersymmetric standard model (MSSM) one introduces a discrete
symmetry, called R-parity, to prevent too fast proton decay. This
symmetry also ensures consistency with electroweak precision
measurements. This discrete symmetry then guarantees the stability
of the lightest R-parity odd particle, in this case the lightest
supersymmetric particle (LSP). Within this context a number of
studies have examined the impact of the precise measurement of the
relic density and shown that it puts strong constraints on
supersymmetric models,  either the MSSM ~\cite{Profumo:2004at,
Allanach:2004xn, Arkani-Hamed:2006mb} or high scale models such as
minimal supergravity (mSUGRA) ~\cite{Ellis:2003si, Baer:2003yh,
Pallis:2003jc, Chattopadhyay:2003xi,  Baltz:2004aw,
Belanger:2005jk, Djouadi:2006be}, non-universal
SUGRA~\cite{Ellis:2002iu, Bertin:2002sq, Birkedal-Hansen:2002am,
Baer:2005bu, Belanger:2004hk}, or string inspired
models~\cite{Binetruy:2003yf}. Furthermore, studies of relic
density of dark matter in some generalizations of the MSSM such as
the MSSM with CP violation\cite{Balazs:2004ae,
Nihei:2004bc,Choi:2006hi,Belanger:2006qa}, the NMSSM which
contains an extra singlet \cite{Belanger:2005kh,Gunion:2005rw} or
even the MSSM with an extended gauge structure such as an extra
U(1)\cite{Barger:2004bz} or the left-right symmetric
model~\cite{Demir:2006ef}, all emphasize the presence of new
channels that can lead to  a reasonable value of the relic density
of dark matter where it was not possible within the MSSM. However
candidates for dark matter  go far beyond the much studied
neutralino LSP in the MSSM. Besides the extensions of the MSSM
mentionned above, explicit examples include a model with universal
extra dimensions~\cite{Kong:2005hn, Servant:2002aq} where the cold
dark matter (CDM) candidate is the partner of the hypercharge
gauge boson, a model with warped extra
dimensions~\cite{Agashe:2004bm} where the CDM is the Kaluza-Klein
excitation of a right-handed neutrino and little Higgs models
where in some cases the partner of the hypercharge gauge boson is
the CDM candidate~\cite{Birkedal-Hansen:2003mp,Hubisz:2004ft}
while in other cases the CDM is a heavy
neutrino~\cite{Martin:2006ss}. One can even have a scalar CDM
candidate, for example in  theory space models
~\cite{Birkedal-Hansen:2003mp}. In all these models that propose
alternate solutions to the hierarchy problem, a discrete symmetry
like R-parity conservation is either present by construction or
needs to be introduced for the viability of the model. The main
problems that this discrete symmetry alleviates are rapid proton
decay, as in  warped extra dimensions, or consistency with
electroweak precision measurements as in universal extra
dimensions or the littlest Higgs model.
 These discrete symmetries then naturally ensure the stability
of the lightest odd particle (LOP) \footnote{In the following we
will use R-parity to designate generically the discrete symmetry
that guarantees the stability of the LOP.}.

Sophisticated tools have been developed to perform a precise
computation of the relic density within R-parity conserving
supersymmetric models, three are publicly available: \micro
~\cite{Belanger:2001fz,Belanger:2004yn},\darksusy~\cite{Gondolo:2004sc}
and \isatools~\cite{Baer:2002fv}.  In view of the diversity of the
new physics models and of the CDM candidates, there is a clear
need for complete and precise codes to calculate the relic density
of dark matter in different models, to the level of what has been
achieved in the MSSM.  Here we propose an extension of \micro~
that serves this purpose.

One of the difficulty of the computation of the relic density is
that a large number of channels can contribute to the
(co-)annihilation cross-section of dark matter. For example  in
the general MSSM, close to 3000 processes can contribute, in
particular a large number of processes are relevant when the
spectrum is such that many particles are not much heavier than the
dark matter candidate. Therefore a certain level of automation is
desirable. The structure of {{\tt micrOMEGAs}} which is based on
{{\tt CalcHEP}}~\cite{Pukhov:2004ca} a program that automatically
calculates cross-sections in a given model, makes it in  principle
straightforward to make a complete code for the relic density
calculation in a new model. Once the new model is implemented into
\calchep~ and Feynman rules defined, all annihilation and
coannihilation channels are calculated automatically. The standard
\micromegas~ routines can then be used to compute the relic
density. To automatize as much as possible the procedure for
implementing a new model,  it is possible to use a program like
LanHEP\cite{Semenov:2002jw}, which starts from a Lagrangian in a
human readable format and derives all the necessary Feynman rules.
Note that a similar approach has been adopted by the authors of
{\tt DM++} ~\cite{Birkedal:2006fz}. Using {\tt
CalcHEP}~\cite{Pukhov:2004ca} for the automatic generation of
matrix elements, they have computed the relic density of dark
matter in a little Higgs model.

 The approach that we propose here is  very general and,
as long as one sticks to tree-level masses and cross-sections,
necessitates minimal work from the user beyond the definition of
the model file. However it has been demonstrated that for an
accurate relic density calculation is is necessary in many cases
to take into account higher-order corrections. In particular
corrections to the mass of either the LOP or of any particle that
can appear in s-channel are important. In the MSSM or in the NMSSM
such corrections to Higgs masses are known to be very  important.
The large QCD corrections to the Higgs width must also be taken
into account when annihilation occurs near a Higgs resonance. In
general one expects that these  loop corrections can be
implemented via an effective Lagrangian. In practice then it might
be necessary for the user to implement additional routines or
interface other programs to take these effects into account.

Another advantage of our approach based on a generic program like
{{\tt CalcHEP}} is that one can compute in addition any
cross-section or decay width in the new model considered. In
particular, tree-level cross-sections for $2\rightarrow 2$
processes and 2-body decay widths of particles are available.
Furthermore the cross-sections times relative velocity, $\sigma
v$, for neutralino annihilation at $v\rightarrow 0$ and the yields
for the continuum $\gamma,e^+,\bar{p},\nu$ spectra, relevant for
indirect detection of neutralinos, are also automatically
computed. The procedure that derives the $\gamma,e^+,\bar{p},\nu$
spectrum from  the different channels for  two-body annihilation
of the dark matter particles  was developed for the MSSM
~\cite{Brun:2006cj} and can be applied to a generic model.

As compared to an earlier version of \micro~ that was developed
specifically for the MSSM,  \micronew~  gives the possibility to
compute  the relic density in a generic model. Furthermore,
improvements have been made  within the MSSM and its extensions.
Note however that the relic density calculation itself has not
been modified for the MSSM in this new version. Specifically, the
new features of \micronew~ are:

\begin{itemize}
\item{} Possibility to include in the package any particle physics
model with a discrete symmetry that guarantees the stability of
the cold dark matter candidate (LOP) and to compute the relic
density of CDM.
\item{} Compute automatically the cross-sections for annihilation
of the LOP at small velocities into SM final states and provide
the energy spectra for $\gamma,e^+, \bar{p},\nu$ final states.
\item{} For the MSSM with input parameters defined at the GUT scale,
the interface with any of the spectrum calculator codes reads an
input file in the {\it SUSY Les Houches Accord format}
(SLHA)~\cite{Skands:2003cj}.
\item{} Implementation of the MSSM with complex parameters (CPV-MSSM)
with an interface to CPsuperH to calculate the spectrum.
\item{} Routine to calculate the electric dipole moment of the electron in
the CPV-MSSM
\item{} In the NMSSM, new interface compatible with NMHDECAY2.1.
\end{itemize}

Generically \micronew~ was developed under the assumption that the
model contains a symmetry like R-parity and that particles are
either odd or even under this parity. In the case of a model with
a $Z_3$-parity, one can effectively use the classification of odd
and even particles and the model can be incorporated in
\micromegas~ as for the simple R-parity model. The extension to
more complicated symmetry groups, especially when they lead to two
stable particles, is not straightforward. Such models cannot be
implemented through an automatic procedure as described here. We
further assume, as is the case in the MSSM, that odd particles
which contribute to coannihilation processes are in relative
thermal equilibrium in the early universe and that for even
particles produced in annihilation of the LOP, there are no
significant branching ratios into odd particles.

 In this manual,  we first briefly review the relic density
calculation. In section 3 we then present the various features of
\micronew~ and the routines available for handling parameters,
computing the relic density, all cross-sections and decay widths
as well as the annihilation cross-sections at small velocities. A
description of specific routines developed for the MSSM, for the
CPV-MSSM  and for  the NMSSM follows in section 4. We then
describe the implementation of new models in \micronew~ in Section
5. Finally we explain the installation procedure. Examples of a
\micronew~ session and possible problems in compilation  can be
found in the appendices.

\section{Relic density of dark matter}

A relic density calculation entails solving the evolution equation
for the abundance of the dark matter, $Y(T)$, defined as the
number density divided by the entropy density, (here we follow
closely the approach in \cite{Gelmini:1990je,Edsjo:1997bg})
\begin{equation}
 \frac{dY}{dT}= \sqrt{\frac{\pi  g_*(T) }{45}} M_p <\sigma v>(Y(T)^2-Y_{eq}(T)^2)
    \label{dydt}
\end{equation}
where $g_{*}$ is an effective number of degree of freedom
\cite{Gelmini:1990je}, $M_p$ is the Planck mass and $Y_{eq}(T)$
the thermal equilibrium abundance. $<\sigma v>$ is the
relativistic thermally averaged annihilation cross-section. The
dependence on the specific model for particle physics enters only
in this cross-section which includes  all  annihilation and
coannihilation channels,
\begin{equation}
       <\sigma v>=  \frac{ \sum\limits_{i,j}g_i g_j  \int\limits_{(m_i+m_j)^2} ds\sqrt{s}
K_1(\sqrt{s}/T) p_{ij}^2 \sum\limits_{k,l}\sigma_{ij;kl}(s)}
                         {2T\big(\sum\limits_i g_i m_i^2 K_2(m_i/T)\big)^2 }\;,
\label{sigmav}
\end{equation}
where $g_i$ is the number of degree of freedom,  $\sigma_{ij;kl}$
the total cross-section for annihilation of a pair of
supersymmetric particles with masses $m_i$, $m_j$ into some
Standard Model particles $(k,l)$, and  $p_{ij}(\sqrt{s})$ is the
momentum (total energy) of the incoming particles in their
center-of-mass frame.

 Integrating Eq.~\ref{dydt} from $T=\infty$ to
$T=T_0$  leads to the present day  abundance $Y(T_0)$
 needed in the estimation of the relic density,
\begin{equation} \label{omegah} \Omega_{LOP} h^2= \frac{8 \pi}{3}
\frac{s(T_0)}{M_p^2 (100{\rm(km/s/Mpc)})^2} M_{LOP}Y(T_0)=
 2.742 \times 10^8 \frac{M_{LOP}}{GeV} Y(T_0)
\end{equation}
where $s(T_0)$ is the entropy density at present time and $h$ the
normalized Hubble constant.

To compute the relic density, \micromegas~ solves the equation for
the abundance Eq.~\ref{dydt}, numerically without any
approximation. In addition, \micromegas~ also estimates the
relative contribution of each individual annihilation or
coannihilation channel to the relic density. For this specific
purpose only, we use the freeze-out approximation (for more
details, see ~\cite{Belanger:2006qa}).

As in previous versions, we include in the thermally averaged
cross-section, Eq.~\ref{sigmav}, only the processes involving the
LOP as well as those  particles for which the Boltzmann
suppression factor, $B$, is above some value $B_\epsilon$

 \beqn
 \label{beps}B=\frac{K_1((m_i+m_j)/T)}{K_1(2m_{LOP}/T)}\approx
e^{-X\frac{(m_i+m_j-2M_{LOP})}{M_{LOP}}}
>B_\epsilon \eeqn
where $m_i,m_j$ are the masses of the incoming particles,
$X=M_{LOP}/T$. The value recommended is
$B_\epsilon=10^{-6}$~\cite{Belanger:2001fz} and corresponds
roughly to $m_{\chi_i} < 1.5 M_{LOP}$. As in previous versions of
{{\tt micrOMEGAs}~, new processes are compiled and added only when
necessary, in run-time.

In the framework of the MSSM, as realized in \micro~,  the
computation of all annihilation and coannihilation cross-sections
are done exactly at tree-level. For this we rely  on {{\tt
CalcHEP}}~\cite{Pukhov:2004ca}, a generic program which once given
a model file containing the list of particles, their masses and
the associated Feynman rules describing their interactions,
computes any cross-section in the model. To generalize this
program to other particle physics models one only needs to replace
the calculation of the thermally averaged annihilation
cross-section for the stable particle that plays the role of dark
matter. This can be done easily after specifying the new model
file into {{\tt CalcHEP}}. Then to solve numerically the evolution
equation (Eq. \ref{dydt})and calculate $\Omega h^2$ one uses the
standard \micromegas~ routines. Full details on this can be found
in Ref.~\cite{Belanger:2004yn}

In order that the program finds the list of processes that need to
be computed for the effective annihilation cross-section, one
needs to specify the analogous of R-parity and assign a parity odd
or even to every  particle in the model. The standard model
particles and other particles such as Higgses have an even parity.
The lightest odd particle will then be identified to the dark
matter candidate.

After specification of the R-parity odd particles, \micromegas~
automatically generate all  processes of the type
$$\sim\chi_i \sim\chi_j \rightarrow X,Y$$ where $\sim\chi_i$ designates all
R-parity odd particle and X,Y all R-parity even particles, for
example standard model particles and Higgses. \micromegas~ then
looks for s-channel poles as well as for thresholds to adapt the
integration routines for higher accuracies in these specific
regions, and performs the relic density calculation. Note that
there is no automatic procedure to check that the LOP is colorless
and neutral,  the relic density calculation can be performed even
for such candidates\footnote{This can be used for example to
calculate the density of the charged NLSP from which one can
extract the relic density of CDM in models where the gravitino is
the LSP ~\cite{Feng:2003uy}.}.

\section{Structure and general routines}
\label{general}

The
 \verb|micromegas_2.0| package contains the  following files and directories: \\
$\bullet$  the directory \verb|CalcHEP_src|  with \calchep~ source files;\\
$\bullet$  the directory  \verb|sources| which  contains general
routines   for relic density  calculation;\\
$\bullet$  the directory MSSM with the MSSM  model files and
auxiliary routines necessary for the model implementation as well as constraints on the model;\\
 $\bullet$ the \verb|Makefile|  used for
package installation.\\
$\bullet$ the \verb|newProject| command file  for
implementation of new models.\\
$\bullet$ the \verb|cgwRun| command file to run the program under
Cygwin.\\

One can also include additional packages developed for specific
models. These packages contain a directory with the name of the
model, e.g. NMSSM or CPV-MSSM, and they should be installed
separately in the main \verb|micromegas_2.0| directory.

The directory for each model, including the new model that can be
defined by the user,
all have the same  structure, they contain three directories\\
\verb|  calchep/   work/  lib/ |\\
\noindent as well as  a \verb|Makefile|,  that compiles the code,
and some sample main programs for the calculation of $\Omega h^2$.
In general, \verb|work| is intended for model implementation and
\calchep~sessions needed for the generation of matrix elements
\verb|lib| - for all special functions necessary for the model and
\verb|calchep| - for interactive \calchep~ sessions only.

The  \verb|work/| directory contains several files and
directories, in particular  the \verb| models/| subdirectory where
the model is defined in \calchep~ notation. The contents of this
subdirectory will be detailed in Section ~\ref{NewModel}, at this
point we only need to mention that two files \verb|vars1.mdl| and
\verb|func1.mdl| contain the independent parameters and functions
of the model. The \verb|work/so_generated| subdirectory is used to
store shared libraries of matrix elements generated by \calchep~.
 When one \micromegas~
routine needs a matrix element it looks for the corresponding
dynamic library in the \verb|so_generated/| directory. If this
library exists it is linked, if not, \micromegas~ launches a
\calchep~ session in \verb|work/calchep| and generates the
requested library. Note however that the specific contents of the
library are not checked, when modifying a model it is necessary to
remove all files in \verb|so_generated/|.

In this section we review the general routines of \micronew~ that
can be used in any model. These routines are located in the
\verb|sources| directory and  compiled in
\verb|sources/micromegas.a|. Most of these routines were described
in more details in Ref.~\cite{Belanger:2004yn} in the context of
the MSSM. We include them here for the sake of completeness. First
note that we provide two sets of programs and routines to allow
the user to work either with {\it C} or {\it Fortran}. Most of the
{\it C} and {\it Fortran} commands have the same format. We list
first the {\it C} commands and when necessary
 the analogous {\it Fortran} call in  squared
 brackets. We do not write explicitly the Fortran call when it is
 identical to the C call \footnote{The \& symbol is
 used in C to designate the address of the corresponding
 parameter, it is not needed in Fortran.}.
  Note that in C, the variable {\it file}
is of {\it FILE*} type, whereas in Fortran it is an integer which
specifies the input/output channel number. In both cases the {\it
fname} variable which specifies the file is used for a text type
variable. All information about the variable types  can be found
in the files \verb|sources/micromegas.h| (C version) and
\verb|sources/micromegas_f.h| (Fortran version). Finally, many of
the functions described in this section are used in the sample
{\it main} routines provided with the package and described in
Section~\ref{mssm}. It could be instructive for the user to study
the examples included with the MSSM and/or the NMSSM.

\subsection{Model parameters}

The parameters needed for the computation of cross-sections in a
given model are specified within \micronew~ in the notation of
\calchep.  These include both the independent parameters of the
model, hereafter also called variables, as well as all internal
functions, the so-called constraints of the models. All model
files can be found in  the \verb|work/models| subdirectory of each
model. The variables are defined in the {\tt vars1.mdl} model file
while the constraints are derived from the independent parameters
and are calculated by \calchep~ as specified in the {\tt
func1.mdl} model file. Both these files contain some comments to
explain the meaning of parameters. Example of constraints are the
vertex functions written in terms of independent parameters of the
model. In some models, masses are constraints whereas in the MSSM,
masses are independent variables. The variables used in the MSSM
in fact include all masses and mixing matrices, a partial list
 can be found in Ref.~\cite{Belanger:2004yn}, Table 2.
 Note that what we call here
independent parameters from the point of view of computing matrix
elements are not necessarily free parameters of the model. For
example in the MSSM, the independent parameters can be derived
from a much reduced set of input parameters, the 5 input
parameters of the SUGRA model or the soft SUSY breaking parameters
of the MSSM. Within \micromegas~  there is no check  of the
self-consistency of the model, we assume that constraints are
imposed by other functions. Therefore one has to be careful when
changing by hand some of the variables of the model. For example,
the neutralino mass matrix depends at tree-level on only four
parameters, so changing the value of the mass of  only one
neutralino without appropriate modifications to other masses and
to the mixing matrix to ensure a self-consistent system could lead
to wrong results.

\subsection{Setting of  parameters.}

In the following we describe the few routines that allow to set or
read the value of any independent
parameters  listed in the {\tt vars1.mdl} file.\\

\noindent
 $\bullet$ \verb|assignVal(name,val)| and
 \verb|assignValW(name,val)| assigns value {\it val} to variable {\it name}.\\
$\bullet$ \verb|findVal(name,&val)| finds the  value of
 variable  {\it name} and assigns it to parameter {\it val}.\\
$\bullet$ \verb|findValW(name)| just returns the value of variable
{\it name}.

\noindent
 If {\it name} does not correspond to any variable of
the model, both  \verb|assignVal| and \verb|findVal| return a
non-zero error code whereas
\verb|assignValW| and \verb|findValW|  write a warning on the screen.\\
\noindent
 $\bullet$ \verb|readVar(file)| reads variables from the
\verb|file|. This file should contain two columns, the first one
specifying the names of variables, and the second one the
corresponding numerical values.   \verb|readVar| returns zero when
the file has been read successfully, a negative value when the
file can not be opened for reading and  a positive  value to
signal a wrong file record at the line corresponding to the value
of the error code.\\
$\bullet$ \verb|printVar(file)| prints the numerical values of all
variables into  {\it file}. Of course, this file should have been
opened previously.\\
 $\bullet$\verb|findParam(name,&err)| returns the value
 of the  parameters {\tt name}  defined in {\tt  models/func1.mdl}.
Not all parameters are accessible automatically by this command,
for details see Section~\ref{NewModel}. A non-zero error code
means that a parameter was not found.

\subsection{ Calculation of Relic density.}
\label{relic}

The first step before performing a relic density calculation is to
find among all the R-parity odd particles, the lightest stable
one. Once this is done, the generic \micromegas~ routine for
calculating the relic density can be called. The routines that
find the LOP, give information about particle masses and calculate
the relic density are described below.\\

\noindent
 $\bullet$ \verb|sortOddParticles(message)|  sorts the odd
particles with increasing  masses. {\it message} contains the name
of the LOP. This routine returns a non zero error code when a
wrong set of parameters is used, for example one for which some
constraints cannot be calculated, the corresponding constraint is
listed in \verb|message|. This routine has to be called before any
other routine described in this section.

\noindent
 $\bullet$ \verb|lopmass_()[lopmass()]|  gives the
mass of the lightest odd particle.

\noindent
 $\bullet$ \verb|printMasses(file,sort)| prints all
masses of odd particles  into the file {\tt file}. If $sort\ne 0$
the masses are sorted so the mass of the LOP is given first.

\noindent $\bullet$ \verb|darkOmega(&Xf,fast,Beps)| is the main
routine to calculate $\Omega h^2$.  $X_f=M_{LOP}/T_{f}$
characterizes the freeze-out temperature.  This routine does not
use the freeze-out approximation. The value of $X_f$ is given for
information and is also used as an input for the routine that
gives the relative contribution of each channel to $\Omega h^2$,
see \verb|printChannels|  below. The  $fast=1$ flag forces the
fast calculation (for more details see
Ref.~\cite{Belanger:2004yn}). This is the recommended option and
gives an accuracy around $1\%$. The parameter $Beps$ defines the
criteria for including a given channel in the computation of the
thermally averaged cross-section, Eq.~\ref{beps}.   The
recommended value is $Beps=10^{-4} - 10^{-6}$, on the other hand
if $Beps=1$ only annihilation of the
lightest odd particle is computed.\\
 $\bullet$
\verb|printChannels(Xf,cut,Beps,prcnt,file)| writes into the file
{\tt file} the  contributions  of different channels to $(\Omega
h^2)^{-1}$. The $cut$ parameter specifies  the lowest value to be
printed. If $prcnt\ne 0$ the contributions are given in percent.
Note that it is only for this specific purpose that we use the
freeze-out approximation.

\subsection{ Calculation of cross-sections and widths.}

For a relic density calculation one needs to compute the
cross-sections for annihilations of any  pairs of R-parity odd
particles. In \micromegas~ the codes for the generation of the
corresponding matrix elements are generated using \calchep. In
fact any $2\rightarrow 2$ cross-section or two-body decay width
within a given model is also available in \calchep, for example
cross-sections for production of pairs of SUSY particles in
$e^+e^-$ collisions. Cross-sections involving 3 or more particles
in the final state can also be computed by \calchep~ but since
special care is required for the phase space integration, for
example  in case of resonances, they are not provided
automatically. To obtain a given cross-section or decay width, the
first step consists in generating  the corresponding matrix
element with \calchep. Then one can check the contents of the new
libraries and finally perform the numerical calculation of
cross-sections or decay widths.\\

\noindent
$\bullet$\verb| newProcess(procName, libName) [ newProcess(procName, libName, address) ]|\\
compiles the  codes for any $2\ra 2$ or  $1\ra 2$  reaction.
The result of the compilation is stored in the library\\
\hspace*{3cm} \verb|work/so-generated/|{\it libName}\verb|.so|.\\
 If the library {\tt libName} already exists, it is not recompiled and the correspondence
between the contents of the library and the {\it procName}
parameter is not checked. {\it libName} is also inserted into the
names of routines in the {\it libName}.so library.  Thus {\it
libName} can not  contain symbols that cannot be used in
identifiers, for example the symbols $+,-,*,/,~$.
The name of a given process, {\it procName},  has to  be specified in CalcHEP notation, for example  in the MSSM\\
\hspace*{3cm} \verb| "e,E->~1+,~1-"|\\
stands for the lightest chargino pair production in $e^+e^-$
collisions. Note that {\it procName} should not contain any blank
space. Multi-process generation is also possible by using the
symbol {\tt 2*x}. For example,  \verb| "e,E->2*x"| designates all
possible two particle final states for an $e^+e^-$ collision. Note
that all library names starting with \verb|omg| or \verb|2width_|
are reserved for internal calls of the {\tt darkOmega} routines
and cannot be used for new libraries. Although such libraries
cannot be created by the user, the ones already compiled in
\micromegas~ can be loaded and used to calculate the corresponding
matrix elements. In this case the {\it procName} argument can be
left blank. These internal \micromegas~ libraries are named
\verb|omg<particle>_<particle>.so | and
\verb|2width_<particle>.so|. Here  \verb|<particle>| is the
particle name where "+", "-", \verb|"~"| are replaced respectively
by "\_P", "\_M", "\_t". The \verb|newProcess| routine returns the
{\it address} of the compiled code for further usage.   If the
process can not be compiled, then a NULL address is
returned\footnote{ In Fortran, instead of  {\it address} we use a
two element  {\tt INTEGER} array, this length is sufficient to
store a computer address.}.\\

There are two routines which allow to check the library contents.\\

\noindent
$\bullet$\verb|procInfo1(address,&ntot,&nin,&nout)|\\
provides information  about the total number of subprocesses
(ntot) stored in the library  specified by {\tt address} as well
as the number of incoming (nin) and outgoing (nout) particles for
these
subprocesses. Typically, for collisions (decays), $nin=2(1)$ and $nout=2$.\\
$\bullet$\verb|procInfo2(address,nsub,N,M)|\\
fills for subprocess $nsub$ ($1\leq nsub \leq ntot$) an array of
particle names $N$ and an array of particle  masses $M$. These
arrays have size $nin+nout$ and the elements are numbered in the
usual CalcHEP
notation starting with the initial state.\\

Once the source code for the relevant matrix elements have been
generated,  two different procedures for the  numerical
calculation of cross-sections or decay widths are available.

\noindent
$\bullet$\verb| cs22(address, nsub, P, c1, c2 , &err)|\\
calculates  the cross-section for a given $2\rightarrow 2$
process, $nsub$, with  center of mass momentum $P$(GeV). The
differential cross-section is integrated
 from  $ c1 < \cos\theta <c2 $  and $\theta$ is
the angle between $\bar{p}_1$ and $\bar{p}_3$  in the
center-of-mass frame. Here $\bar{p}_1$ ($\bar{p}_3$) denote
respectively the momentum of the first initial(final) particle.
{\it err} contains a non zero error code if {\it nsub} exceeds the
maximum value  for the number of subprocesses (given by the
argument ntot in the routine {\tt procInfo1}).

\noindent
$\bullet$\verb| pWidth2(address,nsub)|\\
returns  the partial decay width (in GeV) for subprocess number
$nsub
> 0$. If the parameter $Q$ which
specifies the QCD scale is involved in the width calculation, its
value is automatically set  to the mass of the incoming particle.
For example $Q$   allows to account for running Yukawa coupling in
$h_i\bar{q} q$ vertices.\\
\noindent
$\bullet$\verb| decay2Info(Particle,file)|\\
returns the total width for the particle specified by its name in
\calchep~ notation and writes all partial widths in {\it file}. If
$file=NULL$ (or $file.eq.0$ in Fortran) only the total width is
given as output. This procedure uses the routine \verb|pWidth2|
described above. Note that all the  widths of {\it odd} particles
are independent parameters in the model, see Section
~\ref{specific}. So, for reactions where such widths can play a
role they should first be calculated by \verb|decay2Info|. The
numerical value can be assigned to the appropriate variable via
the \verb|assignVal| command.

Examples on how to use the routines described in this section can
be found in the sample main programs \verb|MSSM/cs_br.|{\it c/F}.

\subsection{ Calculation of annihilation spectra.}

The indirect detection rate of dark matter in the galactic halo
through their decay products in photons, positrons or antiprotons
depend on the annihilation cross-section of dark matter at small
relative velocity. The cross-sections for the different 2-body
annihilation channel of the LOP are calculated automatically in
any model. We then provide the continuum spectrum for $\gamma$,
$e^+$, $\bar{p}$, $\nu$ production. Here we describe the general
procedure used to calculate the spectrum and the different
routines available. Note however that an improved and more
complete version of the indirect detection module, including
integration over various  dark matter profiles and propagation of
positrons and antiprotons will be presented in a separate
publication ~\cite{Brun:2006cj, Belanger:prep}.

The procedure we follow is similar to the one implemented in {\tt
DarkSUSY}~\cite{Gondolo:2004sc}. For the basic channels,
$q\bar{q},\mu^+\mu^-,\tau^+\tau^-,W^+W^-,ZZ$, we provide tables
for $\gamma$, $e^+$, $\bar{p},\nu$ production as obtained by
PYTHIA \footnote{For this version, we use the {\tt DarkSUSY}
tables. Improved and more versatile tables will be presented in
~\cite{Brun:2006cj}.}. For channels containing two different
particles, $AB$, we obtain the final spectrum by taking half the
sum of the $A\bar{A}$ and $B\bar{B}$ spectra.
 For channels with Higgses, or other particles whose mass are a priori unknown,
we recursively calculate all $1\to 2$ decay channels until we
obtain particles in the basic channels. If during these decays we
get a pair of particles $AB$ where $A$ is one of the basic
channel, we suppose that half of the spectrum is obtained from
$A\bar{A}$  and continue to decay $B$.

The gamma ray flux can be evaluated as
\begin{equation}
   \Phi_\gamma = \frac{\sigma v}{M^2_{LOP}} N_{tot}  H\;\; \left(\frac{photons}{cm^2
sec\; sr}\right) \label{flux}
\end{equation}
where $N_{tot}$ is the  number of particles with energy $E
>E_{min}$. The factor $H$ includes the integral of the squared of the dark
matter density over the line of sight,
\begin{equation}
H = \frac{1}{8\pi}\int_0^{\infty} dr
\overline{\rho_Q^2}\left(\sqrt{r^2+r_0^2-2rr_0\cos(\phi)}\right)
\label{halo}
\end{equation}
where $\phi$ is the angle in the direction of observation, in
radians, $\overline{\rho_Q^2}(r)$ is the averaged squared dark
matter density in  $\rm{GeV/cm}^3$, $r$ is the distance  from the
Galactic center in kpc and $r_0=8.5$kpc is the  distance of the
Sun to the center of the galaxy. In the present version, we have
implemented only the modified isothermal distribution,
\begin{equation}
\overline{\rho_Q^2}(r) = \left(\frac{0.3GeV}{cm^3}
\frac{1+(r_0/a_0)^2}{1+(r/a_0)^2}\right)^2 \label{rhoq}
\end{equation}
where $a_0= 3.5$kpc is the length scale.

The main routines to calculate the  $\gamma, e^+,\bar{p},\nu$ spectrum are \\
$\bullet$ \verb|calcSpectrum(v,outP,tab,&err)|\\
calculates $\sigma v$ in $cm^3/sec$ and writes the spectrum of one
collision in the array $tab$. This array  has to contain 250
elements of type $double$.  The input parameters are the relative
velocity $v$ in natural units,\footnote{ Note that for neutralino
collisions in the galactic halo $v$ should be about 0.001.} and
the type of the outgoing particle, $outP$. We use $outP=0,1,2$ for
-$\gamma$, $e^+$, $\bar{p}$ and  3, 4, 5 for $\nu_e$, $\nu_\mu$,
$\nu_\tau$. A non-zero error code indicates that one of the
particle appearing in the
decay products does not have 2-body decay modes.\\
$\bullet$\verb|zInterp(x,tab)|\\
interpolates the  table \verb|tab| obtained by
\verb|calcSpectrum|. Here $x=log(E/M)$, where E is the energy of
the outgoing particle and M the LOP mass. This returns
the value   $dN/dx$, where N is the number of particles. \\
$\bullet$ \verb|spectrInfo(Xmin,tab,&Ntot,&Etot)| \\
calculates some statistical information about the spectra  stored
in table \verb|tab|.  $Xmin = E_{min}/M_{LOP}$ defines the minimal
energy considered.  The routine calculates $Ntot$ - the  number of
particles with energy $E > E_{min}$; $Etot$ - the total energy
(divided by $M_{LOP}$) of the particles produced. When working in
C, NULL  can be substituted as an argument for any unnecessary
output parameters.\\
$\bullet$\verb|spectrTable(tab,fname,mess,Xmin,N)| writes in file
{\it fname} the spectrum stored in the {\it tab} array. {\it mess}
contains some  text that describes the plot. {\it Xmin} is the
minimal energy, $N<300$ is the number of points. This file can be
read by the \calchep~ program \verb|tab_view| (see example in
spectrum.c file) or by other graphics program.\\
$\bullet$\verb|rhoQisothermal(r)| is the averaged squared dark
matter density with the  modified isothermal distribution,
Eq.~\ref{rhoq}, \verb|r| is the
distance from the galactic center in kpc.\\
$\bullet$\verb|HaloFactor(fi, rhoQ)| performs the integration of
the squared dark matter density over the line of sight,
Eq.~\ref{halo}. The function \verb|rhoQ|  can be either
the default function \verb|rhoQisothermal| or another
function  provided by the user.\\

\subsection{QCD routines.}

Many auxiliary routines were developed within the context of a
specific model and will be detailed in the next section. However a
few functions are available for all models, for example the
functions that  compute the running standard parameters: QCD
coupling and heavy quark masses.

\noindent$\bullet$\verb|initQCD(alfsMZ,McMc,MbMb,Mtp)|\\
This function initializes the parameters needed for the functions
listed below. It has to be called before any of these functions.
The input parameters are the QCD coupling at the Z scale,
$\alpha_s(M_Z)$, the quark masses, $m_c(m_c), m_b(m_b)$ and
$m_t(pole)$.

\noindent$\bullet$\verb| alphaQCD(Q)|\\
calculates the  running $\alpha_s$ at the scale \verb|Q| in the
$\overline{MS}$ scheme. The calculation is done using the
\verb|NNLO| formula in \cite{Eidelman:2004wy}. Thresholds for
b-quark and t-quark  are included in  $n_f$ at the scales $\mbmb)$
and $\mt(\mt)$ respectively.

\noindent$\bullet$\verb| MtRun(Q), MbRun(Q), McRun(Q) | \\
calculates top, bottom and charm quarks running masses evaluated
at NNLO.

\noindent$\bullet$\verb| MtEff(Q), MbEff(Q), McEff(Q),  | \\
calculates effective top, bottom and charm quark masses using
~\cite{Eidelman:2004wy}
\begin{eqnarray}
\label{meff}
 M_{eff}^2(Q)&=&M(Q)^2\left[1+5.67a + (35.94-1.36n_f)a^2 \right.\nonumber\\
 &+& \left.(164.14-n_f(25.77-0.259n_f))a^3\right]
\end{eqnarray}
where $a=\alpha_s(Q)/\pi$,    $M(Q)$  and $\alpha_s(Q)$    are the
quark masses and running strong coupling  in the
$\overline{MS}$-scheme. In \micromegas~, we use the effective
quark masses calculated at the scale $Q=2 M_{LOP}$.

\section{Sample models : specific routines}

\subsection{MSSM}\label{mssm}

In the case of the MSSM, special routines were developed both for
specifying the independent parameters as well as for including
higher order corrections. These have already been described in
\cite{Belanger:2004yn}, for completeness we summarize the main
points here and we point out the modifications implemented. In the
MSSM, we use loop corrected superparticle masses and mixing
matrices. These masses and mixing matrices are then used to
compute exactly at tree-level all annihilation/coannihilation
cross-sections. Higher order corrections to the Higgs masses are
also calculated by one of the spectrum calculators. QCD
corrections to Higgs partial widths are included as well as the
important SUSY corrections, the $\dMb$ correction, that are
relevant at large $\tan\beta$. These higher-order corrections also
affect directly the Higgs-$q\overline{q}$ vertices and are taken
into account in all the relevant annihilation cross-sections.
These routines were described in \cite{Belanger:2004yn} and have
not been changed. A description of parameters can be found in the
\verb|MSSM/lib/pmodel.h| file.

The independent parameters of the model include all masses and
mixing matrices as specified in the SLHA~\cite{Skands:2003cj}. We
have chosen this enlarged set of parameters rather than the MSSM
soft SUSY breaking parameters used in the original version of
\micromegas~ for the greater flexibility it provides in modifying
parameters and in incorporating models. In this approach once the
MSSM has been implemented within \calchep, the same model file can
be used for either mSUGRA, AMSB or the general MSSM. The only
difference will be in the definition of the input parameters and
the use of different routines to determine the independent
parameters.

To define the spectrum one can either  read the independent
parameters defined in the {\it SLHA} format \cite{Skands:2003cj}
from an input file,  calculate the set of independent parameters
starting from the GUT scale input parameters and using one of the
spectrum calculator, or calculate the set of independent
parameters from the weak scale MSSM parameters. For input
parameters specified at the GUT scale, for example in the context
of SUGRA models,  loop corrections are obtained from one of the
public codes which calculate the supersymmetric spectrum using
renormalization group equations (RGE), \suspect
\cite{Djouadi:2002ze}, \softsusy \cite{Allanach:2001kg}, \spheno
\cite{Porod:2003um} or \isajet \cite{Paige:2003mg}.

The routines that define the spectrum are

 \noindent $\bullet$ \verb|specSUGRA| defines the
independent parameters of the MSSM starting from a set of input
parameters in the SUGRA model. Here {\tt spec} stands for one of
the spectrum calculators {\tt suspect}, {\tt isajet},
{\tt spheno}, or {\tt softSusy}.\\
\noindent $\bullet$ \verb|specAMSB| does  the same as above within
the {\tt AMSB} model.\\
\noindent
 \noindent $\bullet$
{\tt specEwsbMSSM} calculates the  masses of Higgs  and
supersymmetric particles in the MSSM including one-loop
corrections starting from weak scale input parameters. Here {\tt
spec} stands for one of the spectrum calculators {\tt suspect},
{\tt isajet}, {\tt spheno}, or {\tt softSusy}.\\
\noindent $\bullet$ \verb|assignValW("dMb", deltaMb())| needs to
be called in order to include the threshold correction to the
{$Hb\bar{b}$} vertex. This call  must be done only after the
spectrum has been calculated using one of the commands above.

 Some facilities to read or write directly SLHA files are also
available\\ \noindent$\bullet$ \verb|readLesH(f)| reads the SLHA
input file
{\it f}. \\
$\bullet$ \verb|writeLesH(f)| saves into the file {\it f} the SLHA
MSSM parameters.\\

The  default spectrum calculator package is {\suspect}. To work
with another package one has to specify the appropriate path in
\verb|MSSM/lib/Makefile|. For this  the environment variables
\verb|ISAJET|, \verb|SPHENO| or \verb|SOFTSUSY| must be redefined
accordingly. Note that we also provide a special interface for
\verb|ISAJET| to read a SLHA file. This means that the user must
upgrade his original {\tt ISAJET} libraries to include this
interface.
Specific instructions are provided in the \verb|README| file.\\

There are two spectrum information  commands which in
\micro~ were included in \verb|printMasses|, \\

\noindent $\bullet$  \verb|HiggsMasses(file)|   prints into
\verb|file| the  masses and widths of Higgs particles.\\
\noindent $\bullet$  \verb|o1Contents(file)|  prints into
\verb|file| the neutralino LSP components in terms of {\it bino,
wino, higgsino1,} and {\it higgsino2} fractions.

The MSSM package also includes routines that calculate other
constraints such as $\delta\rho$  ( \verb|deltarho_|), $(g-2)_\mu$
(\verb|gmuon_|), collider  limits (\verb|masslimits_|),
$Br(b\rightarrow s\gamma)$ (\verb|bsgnlo_|), and
$Br(B_s\rightarrow \mu^+\mu^-)$ (\verb|bsmumu_|), see
Ref.~\cite{Belanger:2004yn}.


\subsubsection{Sample main programs}

The directory MSSM contains several examples of {\it main}
programs written in C (or  Fortran), these include programs to
calculate $\Omega h^2$ with either SUGRA or MSSM models as well as
programs to calculate cross-sections and branching ratios or
spectrum of outgoing particle for dark matter annihilation. In
general when these sample programs are launched
without arguments, the  arguments  needed are explained on the screen. \\

\noindent
 $\bullet$\verb|sugomg.|c/F  calculates the spectrum of
Higgs and SUSY particles
as well as the relic density $\Omega h^2$ and other constraints.\\
\noindent $\bullet$\verb|s_cycle.|c calculates the relic density
of $\Omega h^2$ in a cycle. For the set of mSUGRA parameters used
in
this example,  the  output should correspond to the file \verb|data/s_cycle.res|.\\
$\bullet$\verb|omg.|c/F calculate $\Omega h^2$ for a set of MSSM
input parameters to be read from a file. The name of this file is
given as an argument  of the executable \verb|omg|. The
appropriate format is given in the sample files provided in the
directory \verb|data|. It is possible to execute the program with
several files (arguments) in one call. The results corresponding
to the sample input data files \verb|  omg data/data*|
 can be found in  the \verb|data/omg.res| file.\\
$\bullet$\verb|cs_br.|c/F  give some examples  for the calculation
of  partial widths and cross-sections.\\
$\bullet$\verb|spectrum.|c/F give examples of calculation of
neutralino annihilation spectrum. The output is a graphical plot
which is displayed on the screen.

\subsection{CPV-MSSM}

In the MSSM model described above all parameters are assumed to be
real, generically though the parameters of the MSSM can be complex
(CPV-MSSM). Furthermore, it is possible that electroweak
baryogenesis could work in MSSM scenarios with complex phases. A
new CPV-MSSM model file with complex parameters was rebuilt in the
\calchep~ notation ~\cite{Pukhov:2004ca} using
\lanhep~\cite{Semenov:2002jw}, thus specifying all relevant
Feynman rules. For the Higgs sector, an effective potential was
written in order to include in a consistent manner higher-order
effects ~\cite{Lee:2003nt}. Although this model is only a simple
extension of the MSSM, the implementation within \micromegas~ is
done differently. The independent parameters of the model include
in addition to some standard model parameters only the weak scale
MSSM input parameters. The constrained parameters, in particular
the masses, are evaluated through auxiliary functions. Masses,
mixing matrices and parameters of the effective potential are read
directly from CPsuperH~\cite{Lee:2003nt}, together with masses and
mixing matrices of neutralinos, charginos and third generation
sfermions.  On the other hand, masses of the first two generations
of sfermions are evaluated (at tree-level) within \micromegas~ in
terms of independent parameters of the model. This means that in
this model it is not possible to get the value of the mass of the
SUSY particles, say the lightest neutralino, using the
\verb|findVal("MNE1")| command, rather one has to call
\verb|findParam("MNE1",err)|.  The code for CPsuperH is included
in our package.

 Among the special routines that were described for the MSSM in
 the previous section, only the ones that give information about the spectrum
 can also be used in this model, \verb|HiggsMasses(file)| and
 \verb|o1Contents(file)|. The latter gives the real and imaginary
 part of the $bino/wino/Higgsino_1/Higgsino_2$ components of the LSP.
Since the phases in the CPV-MSSM are strongly constrained by
measurements of electric dipole moments, it is important ti take
into account. We have developed a special routine for this purpose

\noindent $\bullet$ \verb|edm_(&de,&dTl)| returns the value of the
electric dipole moment of the electron, $d_e$  in units of $ecm$
as well as the dipole moment of Thallium, $d_{Tl}$. One-loop
neutralino/chargino contributions as well as two-loop squark,
quark and chargino contributions are included
~\cite{Belanger:2006qa,Choi:2004rf,Chang:1998uc,Pilaftsis:2002fe}
as well as the four-fermion operator for $d_{Tl}$. The latter two
contributions can dominate, especially for large values of
$\tan\beta$. The upper limit from the measurement of the electric
dipole moment of the electron is actually $d_e < 2.2 \times
10^{-27}$ecm~\cite{Hagiwara:2002fs}.

\subsection{NMSSM}

The NMSSM is the simplest extension of the MSSM with one extra
singlet, as a result the model contains one additional neutralino
as well as additional scalars. A new model file was implemented
into \calchep~ and as in the MSSM, an improved effective potential
for the Higgs sector was defined. More details on the model can be
found in Ref.~\cite{Belanger:2005kh}.

The independent parameters of the model include, in addition to
some standard model parameters, the weak scale NMSSM input
parameters as defined in the SLHA2~\cite{Allanach:2005kk}.   The
constrained parameters, in particular the masses, are evaluated
through auxiliary functions. For this we call
NMHDECAY~\cite{Ellwanger:2005dv} specifying an input file in the
SLHA2 format (slhainp.dat).  The masses,  mixing matrices are then
read directly from the output file of NMHDECAY
~\cite{Ellwanger:2004xm} written in the SLHA2 format (spectr.dat).
These two SLHA files are available in the main NMSSM directory.
The parameters of the Higgs potential are derived from the
physical masses and mixing matrices of the charged and neutral
Higgses as described in ~\cite{Belanger:2005kh}. Note however that
in NMHDECAY, the masses are computed using the $\overline{DR}$
values of the independent parameters of the Higgs sector,
$\lambda,\kappa,\tan\beta,\mu,A_\lambda,A_\kappa$ and include
additional loop corrections. Some of these parameters, in
particular $\mu,A_\lambda,A_\kappa$ receive large corrections. To
take these corrections into account we have modified the procedure
to extract the parameters of the Higgs potential that used
previously  the input value at weak scale. We have a good
agreement with NMHDECAY for the partial widths of heavy  Higgses
into lighter Higgs particles.

The  \verb|HiggsMasses(file)| and \verb|o1Contents(file)| routines
described in  the MSSM can also be used in this model. The latter
gives the $bino/wino/Higgsino_1/Higgsino_2/singlino$ components of
the LSP. As in the MSSM the function \verb|bsgnlo_()| returns the
value of $Br(b\rightarrow s\gamma)$. This value is calculated
within NMHDECAY~\cite{Ellwanger:2004xm,Ellwanger:2005dv}
\footnote{Note that for the moment, NMHDECAY computes only the
one-loop contribution to $Br(b\rightarrow s\gamma)$.}. Note that
in this model it is not possible to get the value of the mass of
the SUSY particles using the \verb|findVal| command, since these
are constrained parameters. Rather one has to call
\verb|findParam|.

Theoretical and experimental constraints on the model are checked
thoroughly within NMHDECAY, a function has been written  specifically for the NMSSM :\\
$\bullet$\verb|NMHwarn(file)| returns the number of constraints
that are not satisfied as provided by NMHDECAY. Information about
these constraints is stored in \verb|file|. The description of
different experimental (e.g. LEP limits on Higgs masses) and
theoretical constraints on the model can be found in
Ref.~\cite{Ellwanger:2004xm}.

\section{ New models.}
\label{NewModel}

In general, to implement a new model the user only needs to write
the model in the  {{\tt CalcHEP}} format.
The directory for the new model can be created by the\\
\verb|    newProject <NewModel>|\\
command. The new directory \verb|<NewModel>| will contain in
particular, a directory \verb|work|  for the description of the
model and a directory \verb|lib| implementation of external
functions.
 It also contains a directory \verb|calchep|  as well as examples of {\it main} programs
for the calculation of $\Omega h^2$, \verb|omg.c| and
\verb|omg.F|, and finally a \verb|Makefile| that  compiles the
code. Note that this \verb|Makefile| is created automatically and
does not need to be modified by the user.

We have tried to minimize the amount of user ``intervention". In
general the user has only to include the \calchep~ model files in
\verb|work/models|, put the external codes that calculate external
functions of the model as well as auxiliary routines in the
directory \verb|lib|,  write the corresponding \verb|lib/Makefile|
for their compilation. All other files and subdirectories are
generated automatically and do not need to be modified by the
user.

\subsection{The directory work}

 To implement a new model, the first step consists in
writing the new {{\tt CalcHEP}} model files in the sub directory
\verb+work/models+. This model has to be the first  in the list,
thus the  files must have names \verb|*1.mdl|. More precisely the
model must include five files that specify the list of particles
(prtcls1.mdl), the independent variables (vars1.mdl), the
Lagrangian with all vertices (lgrng1.mdl), all internal functions
(func1.mdl) and external libraries required for the model
(extlib1.mdl).
 Note that to automatize as much as possible the procedure for creating
a new model, it is possible to use a program like {{\tt
LanHEP}}\cite{Semenov:2002jw}, which starts from the Lagrangian
and derives all the necessary Feynman rules {\footnote{{{\tt
LanHEP}} was developed for {{\tt CompHEP}}~\cite{Pukhov:1999gg}
but there exists a simple tool to make a conversion to the {{\tt
CalcHEP}} notation.}.} Alternatively the user can write by hand
the model files of the new model.  Slight modifications to the
standard {{\tt CalcHEP}} model files are necessary, specific
requirements are given below in Section~\ref{specific}.

The directory \verb|work/so_generated|  contains the libraries of
matrix elements generated automatically  by \calchep~. As
mentioned earlier, since the contents of libraries in
\verb|so_generated/| are not checked, this directory should be
cleaned every time a model is modified. It will be then
regenerated automatically.

\subsection{The directory lib}

Additional external functions may also be required to have a
complete model, these should be included in the directory
\verb|lib|. The functions required by {{\tt func1.mdl} have to be
incorporated as a shared library {{\tt lib/mLib.so}}\footnote{In
the case of Cygwin the shared libraries have the  extention .dll}.
Other auxiliary routines not needed for the model should be
included in  {{\tt lib/aLib.a}}.

As explained in Section~\ref{mssm}, it is sometimes more
convenient to choose an enlarged set of independent parameters for
a given model even though physically these parameters are not
truly independent. Clearly, constraints on these parameters have
to be imposed before the calculation of matrix elements. Such is
the case in the MSSM implementation described here where all
masses and mixings are chosen as independent parameters. As usual,
independent parameters have to be listed in {{\tt vars1.mdl}} and
should not be  defined in \verb|func1.mdl|. Also the relevant
functions should be compiled in \verb|lib/aLib.a| rather than in
{{\tt lib/mLib.so}}. It is not compulsory to implement one or even
both of these libraries. The presence of these libraries are
always checked by \micromegas~ commands before passing them to the
linker. The execution of  the \verb|Makefile| in {{\tt
<NewModel>}} launches the \verb|lib/Makefile|  in order to update
the user's libraries.

\subsection{Specific requirements for  the implementation of a new model.}
\label{specific}

 The general format to be used for  model files is
described in ~\cite{Pukhov:2004ca}. Here we explain only specific
points needed for \micromegas.

\paragraph{Names of odd particles.}
The name of odd particles must start with \verb|~|. With this
convention,  automatic identification of the R-parity odd
particles is done by \micromegas. For the purpose of optimizing
the code, it is recommended that the first
 odd particle in the {\tt models/prtcls1.mdl} list be a
potential LOP candidate.

\paragraph{Masses of odd particles.}

A \verb|*| symbol should be added before the masses   of R-parity
odd particles that are not independent parameters of the model,
that is the ones that are found  in the \verb|models/func1.mdl|
file. This is to force  the inclusion of the corresponding
parameter in any generated code.

\paragraph{Widths of odd particles.}
The widths of odd particles should be independent parameters of
the model. All odd particles, including the LOP, must have a width
in order to avoid divergences in cross-sections. For example,
 in coannihilation processes with exchange of the LOP in
t-channel, one can meet a pole if the width of the LOP is zero.
The width of the LOP is set within \micromegas~ to $M_{LOP}/100$,
see explanations in \cite{Belanger:2004yn}.

\paragraph{Automatic  calculation of widths.}

In  previous \calchep~ versions the widths of particles were
treated as independent parameters.  Starting from CalcHEP\_2.4
there is an option to calculate widths automatically. To switch on
the mechanism of automatic width calculation one must add the '!'
symbol in front of the width  in the particle  list
(\verb|models/prtcls1.mdl|). If necessary, the corresponding
parameter should be removed from  the list of independent
parameters (\verb|models/vars1.mdl|) and/or from  the list of
constrained parameters (\verb|models/func1.mdl|). For the relic
density calculation, this trick is used for widths of Higgs
particles which occur as s-channel resonances.

\subsection{Compilation of libraries.}

Specific  requirements for writing the \verb|lib/Makefile| are
detailed in this section.  The {\tt mlib.so} library is used for
external names resolution in the generation of {\it shared}
libraries of matrix elements. Such name resolution is a strict
requirement of {\it Darwin} and {\it Cygwin} platforms. We
recommend the use of the compiler flags stored in the
\verb|CalcHEP/FlagsForMake| file, in particular the option {\tt
SHARED} needed to generate a {\it shared library} instead of an
executable code.

Whenever  operating with shared libraries there is a general
 a problem in finding the location  of the library  in run-time.
Normally, each shared library has a record which specifies its
location, while linking this information is passed to the {\it
main} program. Usually in order to write this record correctly one
has to specify the full path of the library after the {{\tt -o}}
instruction. On some other platforms, for instance {\it OSF1}, one
needs an additionnal linker flag. In \verb|flagsForMake|, this
flag is named {{\tt SONAME}} and it should be followed by the full
name of the library.  The \verb|MSSM/lib/Makefile| gives an
example on how to use the  {{\tt SHARED}} and {{\tt SONAME}}
flags. All these options do not work with the {\it Cygwin} version
of UNIX. In this particular case only the shared libraries whose
paths are included in the {{\tt PATH}} environment variable can be
linked in run-time. For this reason,  in the case of {\it Cygwin},
\micromegas~
 executables have to be launched via the command \\
\verb|    ./cgwRun <exec> <param>|\\
which corrects the {\tt PATH} environment parameters before
starting the \micromegas~ executable. Here \verb|exec| stands for
the name of the executable and \verb|param| for the input
parameters.

We assume that {\tt mLib.so} does not contain calls to functions
described in Section~\ref{general} or calls to external functions
not needed for matrix element generation. On the other hand, there
are no such restrictions for {\tt aLib.a}. It can call any
function in \verb|sources/micromegas.a|, for example the {{\tt
assignValW}} function to set the value of independent parameters
of the model. {{\tt aLib.a}}  can also use any function
implemented in {{\tt mLib.so}}, in particular functions used in
\verb|models/func1.mdl|. Thus the user can access the value of
constrained parameters. For example, in the NMSSM,
\verb|nMass(1000022)| will give the value of the mass of the
lightest neutralino. Here the PDG~\cite{Eidelman:2004wy} code is
used for particle names. Alternatively, we provide a more
convenient
function that serves the same purpose. The function \\
\verb|  findParam(name,&err)|\\
will return the value of the parameters \verb|name| defined in
\verb|models/func1.mdl|. A non-zero error code \verb|err|  signals
failure to find the needed parameter. The parameters that are
accessible automatically by
 this command are, for technical reasons, the ones that enter the
 calculation of the matrix element for pair annihilation of the first
 odd particle listed in \verb|prtcls1.mdl|.
 It is possible to access all other parameters in
 {\tt  models/func1.mdl}. For this, the user must insert a * in
 front of the corresponding parameter in {\tt func1.mdl}
 and recompile the model
\footnote{For this, the generated shared libraries for matrix
elements should first be cleaned. It is only  when \micromegas~
recompiles the libraries that  the parameter will be accessible
via {\tt findParam}.}.

\paragraph{ Use of external programs.}
When  one function implemented in the CalcHEP model makes use of
some large external program,
we recommend to use a {\tt system} function call to launch the
corresponding external  program separately. Communication with
this external program can proceed via files. There is no need to
rewrite external functions as shared libraries. For example we
have adopted this procedure in the MSSM  to call spectrum
calculators, the SLHA files provide a format for reading the
input/output files of these external programs. A similar procedure
is used for calls to NMHDECAY or CPSuperH in the NMSSM or the
CPV-MSSM.

\paragraph{Code optimization.}
The evaluation of some external functions or constraints
implemented in the model can be time consuming. Sometimes
\micromegas~ recalculates all constraints several times with
slightly different parameters. We recommend  that the user
inserts, in the time consuming functions, some checks of input
parameters and uses the previous result if the arguments were not
changed.

\subsection{Check of the new model}

We strongly recommend,  before the first launch of \micronew~ with
a new model, to check the model in an interactive \calchep~
session. For this purpose first launch \verb|./calchep| in the
\verb|work| directory, find the \verb|Edit model| menu and make
some modification without really changing the model, say add and
remove one symbol. When you will leave the \verb|Edit model| menu,
after you have  confirmed your corrections, \calchep~ will start
to check the model. You should remove all bugs detected before
starting your \micromegas~ session. The completeness of the {\tt
mLib.so} library for the model can also be checked.
For this one must first add to {\tt work/models/extlib1.mld} one record,\\
\verb|       ../../lib/mLib.so |\\
Then, within an interactive  \calchep~ session, the check will be
performed at the compilation of a new process, for example some
decay width. An error message will signal a problem in the {\tt
mLib.so} library.

\section{Installation}

The package can be obtained from the web page
\verb|wwwlapp.in2p3.fr/micromegas|.
 Unpacking the file
\verb|micromegas_2.0.tgz| will create the directory
\verb|micromegas_2.0| described in previous sections. This file
contains  the full implementation of the MSSM model. The  NMSSM
and the CPV-MSSM model files and auxiliary routines for
implementation of the model are obtained independently from
\verb|NMSSM.tgz| and \verb|CPVMSSM.tgz|. They have to be
downloaded, unpacked, and copied in the \verb|micromegas_2.0|
directory.

The installation consists of two steps, the general installation
of \verb|micromegas_2.0| and the installation of special models.
The general installation  is realized by the
command \\
\verb|    gmake|\footnote{If {\tt gmake} is not available, for
example with Darwin, one should use {\tt make} instead} \\ This
command identifies the Unix  platform, compiles the \calchep~
executable

\noindent \verb|CalcHEP_src/bin/s_calchep| as well as the general
function library \verb|sources/micromegas.a|.

To install a new model one must use the command

\verb|      ./newProject <Name> |\\
which creates a new sub-directory \verb|<Name>|  containing all
files and sub-directories needed for a new model. This directory
has the same structure as the \verb|MSSM| directory but does not
contain special MSSM routines and model files.  How this directory
should be updated to implement completely a new model, was
explained in Section \ref{NewModel}.

At last, in order to compile an executable file for the
calculation of relic density in the framework of a model one has
to move to the
corresponding directory, say \verb|MSSM|, and  call \\
\verb|    gmake  main=<filename>|   \\
where \verb|<filename>| designates the name of C or Fortran {\it
main} routines. This  \verb|<filename>| should have the
corresponding  \verb|.c| or \verb|.F| extension. The  executable
generated will have the same name without an extension. To launch
the executable with \cygwin~ one must instead use the command\\
\verb|    ./cgwRun <name of exec> <param>|\\
 This commands improves the \verb|PATH| environment parameters so
 that the shared libraries are readable.

\section*{Acknowledgements}

This work was supported in part by GDRI-ACPP of CNRS. The work of
A. Semenov was also supported  by grants from the Russian Federal
Agency for Science, NS-8122.2006.2 and RFBR-04-02-17448. The
authors thank A.~Belyaev and U. Ellwanger for fruitful
discussions, S. Kraml for her contributions in the implementation
of the CPV-MSSM and C.~Balazs for the first implementation of
\micronew~ for models with universal extra dimensions. We also
thank C. Hugonie for discussions on NMHDECAY, for his contribution
in the implementation of the NMSSM  and for his help in making a
version compatible with Darwin.

\appendix
\section{Problems in compilation.}
The code was tested on several UNIX platforms with standard
configuration, OSF1, SunOS, Darwin, Cygwin. In general it should
work without special tuning. Some common problems and their
solution are listed below.

At first  \verb|gmake|   launches the \verb|getFlags| routines in
the \verb|CalcHEP_src| subdirectory. This \verb|getFlags| checks
the UNIX platform and  writes the \verb|FlagsForSh| file which
contains compiler flags and special linker options accordingly.
After that \verb|getFlags| checks compiles and writes an error
message, if some option are not available. If \verb|FlagsForSh|
already exists, then \verb|getFlags| only checks compilers. This
way, compiler options can be improved by users.

\section{Example of microMEGAs session.}
Here we present a sample  output of the particle spectrum and
relic density calculation in the case of the mSUGRA model. First
the user has to compile  general routines with \verb|./gmake|
launched in the directory \verb|micromegas_2.0|, then   the C
version of \verb|sugomg| main program should be compiled in the
\verb|MSSM|
directory  with \\
\verb|  gmake main=sugomg.c|\\
This generates the executable \verb|sugomg|. This executable needs
at least 4 parameters, additional parameters are set to their
default value if not provided. The meaning of the input parameters
are written on the screen when launching the programs without
input parameters. Out test run, which uses by default \suspect~
for the spectrum calculation, should give the following output:
\begin{verbatim}
 ./sugomg 100 100 0 10

Higgs masses and widths
Mh    = 100.67 (wh    =2.5E-02)
MHH   = 173.71 (wHh   =8.5E-01)
MH3   = 172.58 (wH3   =1.2E+00)
MHc   = 190.85

Masses of  Odd particles:
~o1 : MNE1  =    28.6 || ~1+ : MC1   =    50.0 || ~o2 : MNE2  =    55.8
~nl : MSnl  =   101.9 || ~ne : MSne  =   102.4 || ~nm : MSnm  =   102.4
~l1 : MSl1  =   107.9 || ~eR : MSeR  =   115.1 || ~mR : MSmR  =   115.1
~eL : MSeL  =   129.6 || ~mL : MSmL  =   129.6 || ~l2 : MSl2  =   134.3
~o3 : MNE3  =   158.0 || ~o4 : MNE4  =   188.2 || ~2+ : MC2   =   190.6
~t1 : MSt1  =   192.5 || ~b1 : MSb1  =   240.6 || ~uR : MSuR  =   254.4
~cR : MScR  =   254.4 || ~uL : MSuL  =   257.4 || ~cL : MScL  =   257.4
~dR : MSdR  =   257.5 || ~sR : MSsR  =   257.5 || ~b2 : MSb2  =   261.0
~dL : MSdL  =   269.4 || ~sL : MSsL  =   269.4 || ~g  : MSG   =   270.1
~t2 : MSt2  =   332.9 ||

~o1 = 0.828*bino -0.302*wino +0.453*higgsino1 -0.135*higgsino2
Omega=     7.01E-02

Channels which contribute to 1/(omega) more than 1%.
Relative contributions in % are displayed
 44% ~o1 ~o1 -> b B
  7% ~o1 ~o1 -> d D
  4% ~o1 ~o1 -> u U
  4% ~o1 ~o1 -> c C
  7% ~o1 ~o1 -> s S
 17% ~o1 ~o1 -> l L
  8% ~o1 ~o1 -> m M
  8% ~o1 ~o1 -> e E
deltartho=   5.64E-04
gmuon=   1.22E-08
bsgnlo=   1.63E-04
bsmumu=   3.74E-09
WARNING: Chargino below LEP limit

\end{verbatim}


\begin{thebibliography}{10}

\bibitem{Bennett:2003bz}
C.~L. Bennett {\em et.~al.}, {\em Astrophys. J. Suppl.} {\bf 148} (2003) 1,
  [\href{http://xxx.lanl.gov/abs/astro-ph/0302207}{{\tt astro-ph/0302207}}].

\bibitem{Spergel:2003cb}
D.~N. Spergel {\em et.~al.},, {\bf WMAP} Collaboration {\em Astrophys. J.
  Suppl.} {\bf 148} (2003) 175,
  [\href{http://xxx.lanl.gov/abs/astro-ph/0302209}{{\tt astro-ph/0302209}}].

\bibitem{Tegmark:2003ud}
M.~Tegmark {\em et.~al.},, {\bf SDSS} Collaboration {\em Phys. Rev.} {\bf D69}
  (2004) 103501, [\href{http://xxx.lanl.gov/abs/astro-ph/0310723}{{\tt
  astro-ph/0310723}}].

\bibitem{Jungman:1995df}
G.~Jungman, M.~Kamionkowski, and K.~Griest, {\em Phys. Rept.} {\bf 267} (1996)
  195--373, [\href{http://xxx.lanl.gov/abs/hep-ph/9506380}{{\tt
  hep-ph/9506380}}].

\bibitem{Profumo:2004at}
S.~Profumo and C.~E. Yaguna, {\em Phys. Rev.} {\bf D70} (2004) 095004,
  [\href{http://xxx.lanl.gov/abs/hep-ph/0407036}{{\tt hep-ph/0407036}}].

\bibitem{Allanach:2004xn}
B.~C. Allanach, G.~Belanger, F.~Boudjema, and A.~Pukhov, {\em JHEP} {\bf 12}
  (2004) 020, [\href{http://xxx.lanl.gov/abs/hep-ph/0410091}{{\tt
  hep-ph/0410091}}].

\bibitem{Arkani-Hamed:2006mb}
N.~Arkani-Hamed, A.~Delgado, and G.~F. Giudice, {\em Nucl. Phys.} {\bf B741}
  (2006) 108--130, [\href{http://xxx.lanl.gov/abs/hep-ph/0601041}{{\tt
  hep-ph/0601041}}].

\bibitem{Ellis:2003si}
J.~R. Ellis, K.~A. Olive, Y.~Santoso, and V.~C. Spanos, {\em Phys. Rev.} {\bf
  D69} (2004) 095004, [\href{http://xxx.lanl.gov/abs/hep-ph/0310356}{{\tt
  hep-ph/0310356}}].

\bibitem{Baer:2003yh}
H.~Baer and C.~Balazs, {\em JCAP} {\bf 0305} (2003) 006,
  [\href{http://xxx.lanl.gov/abs/hep-ph/0303114}{{\tt hep-ph/0303114}}].

\bibitem{Pallis:2003jc}
C.~Pallis and M.~E. Gomez, \href{http://xxx.lanl.gov/abs/hep-ph/0303098}{{\tt
  hep-ph/0303098}}.

\bibitem{Chattopadhyay:2003xi}
U.~Chattopadhyay, A.~Corsetti, and P.~Nath, {\em Phys. Rev.} {\bf D68} (2003)
  035005, [\href{http://xxx.lanl.gov/abs/hep-ph/0303201}{{\tt
  hep-ph/0303201}}].

\bibitem{Baltz:2004aw}
E.~A. Baltz and P.~Gondolo, {\em JHEP} {\bf 10} (2004) 052,
  [\href{http://xxx.lanl.gov/abs/hep-ph/0407039}{{\tt hep-ph/0407039}}].

\bibitem{Belanger:2005jk}
G.~Belanger, S.~Kraml, and A.~Pukhov, {\em Phys. Rev.} {\bf D72} (2005) 015003,
  [\href{http://xxx.lanl.gov/abs/hep-ph/0502079}{{\tt hep-ph/0502079}}].

\bibitem{Djouadi:2006be}
A.~Djouadi, M.~Drees, and J.-L. Kneur, {\em JHEP} {\bf 03} (2006) 033,
  [\href{http://xxx.lanl.gov/abs/hep-ph/0602001}{{\tt hep-ph/0602001}}].

\bibitem{Ellis:2002iu}
J.~R. Ellis, T.~Falk, K.~A. Olive, and Y.~Santoso, {\em Nucl. Phys.} {\bf B652}
  (2003) 259--347, [\href{http://xxx.lanl.gov/abs/hep-ph/0210205}{{\tt
  hep-ph/0210205}}].

\bibitem{Bertin:2002sq}
V.~Bertin, E.~Nezri, and J.~Orloff, {\em JHEP} {\bf 02} (2003) 046,
  [\href{http://xxx.lanl.gov/abs/hep-ph/0210034}{{\tt hep-ph/0210034}}].

\bibitem{Birkedal-Hansen:2002am}
A.~Birkedal-Hansen and B.~D. Nelson, {\em Phys. Rev.} {\bf D67} (2003) 095006,
  [\href{http://xxx.lanl.gov/abs/hep-ph/0211071}{{\tt hep-ph/0211071}}].

\bibitem{Baer:2005bu}
H.~Baer, A.~Mustafayev, S.~Profumo, A.~Belyaev, and X.~Tata, {\em JHEP} {\bf
  07} (2005) 065, [\href{http://xxx.lanl.gov/abs/hep-ph/0504001}{{\tt
  hep-ph/0504001}}].

\bibitem{Belanger:2004hk}
G.~Belanger, F.~Boudjema, A.~Cottrant, A.~Pukhov, and A.~Semenov, {\em Czech.
  J. Phys.} {\bf 55} (2005) B205--B212,
  [\href{http://xxx.lanl.gov/abs/hep-ph/0412309}{{\tt hep-ph/0412309}}].

\bibitem{Binetruy:2003yf}
P.~Binetruy, Y.~Mambrini, and E.~Nezri, {\em Astropart. Phys.} {\bf 22} (2004)
  1--18, [\href{http://xxx.lanl.gov/abs/hep-ph/0312155}{{\tt hep-ph/0312155}}].

\bibitem{Balazs:2004ae}
C.~Balazs, M.~Carena, A.~Menon, D.~E. Morrissey, and C.~E.~M. Wagner, {\em
  Phys. Rev.} {\bf D71} (2005) 075002,
  [\href{http://xxx.lanl.gov/abs/hep-ph/0412264}{{\tt hep-ph/0412264}}].

\bibitem{Nihei:2004bc}
T.~Nihei and M.~Sasagawa, {\em Phys. Rev.} {\bf D70} (2004) 055011,
  [\href{http://xxx.lanl.gov/abs/hep-ph/0404100}{{\tt hep-ph/0404100}}].

\bibitem{Choi:2006hi}
S.~Y. Choi and Y.~G. Kim, {\em Phys. Lett.} {\bf B637} (2006) 27--31,
  [\href{http://xxx.lanl.gov/abs/hep-ph/0602109}{{\tt hep-ph/0602109}}].

\bibitem{Belanger:2006qa}
G.~Belanger, F.~Boudjema, S.~Kraml, A.~Pukhov, and A.~Semenov,
  \href{http://xxx.lanl.gov/abs/hep-ph/0604150}{{\tt hep-ph/0604150}}.

\bibitem{Belanger:2005kh}
G.~Belanger, F.~Boudjema, C.~Hugonie, A.~Pukhov, and A.~Semenov, {\em JCAP}
  {\bf 0509} (2005) 001, [\href{http://xxx.lanl.gov/abs/hep-ph/0505142}{{\tt
  hep-ph/0505142}}].

\bibitem{Gunion:2005rw}
J.~F. Gunion, D.~Hooper, and B.~McElrath,
  \href{http://xxx.lanl.gov/abs/hep-ph/0509024}{{\tt hep-ph/0509024}}.

\bibitem{Barger:2004bz}
V.~Barger, C.~Kao, P.~Langacker, and H.-S. Lee, {\em Phys. Lett.} {\bf B600}
  (2004) 104--115, [\href{http://xxx.lanl.gov/abs/hep-ph/0408120}{{\tt
  hep-ph/0408120}}].

\bibitem{Demir:2006ef}
D.~A. Demir, M.~Frank, and I.~Turan,
  \href{http://xxx.lanl.gov/abs/hep-ph/0604168}{{\tt hep-ph/0604168}}.

\bibitem{Kong:2005hn}
K.~Kong and K.~T. Matchev, \href{http://xxx.lanl.gov/abs/hep-ph/0509119}{{\tt
  hep-ph/0509119}}.

\bibitem{Servant:2002aq}
G.~Servant and T.~M.~P. Tait, {\em Nucl. Phys.} {\bf B650} (2003) 391--419,
  [\href{http://xxx.lanl.gov/abs/hep-ph/0206071}{{\tt hep-ph/0206071}}].

\bibitem{Agashe:2004bm}
K.~Agashe and G.~Servant, {\em JCAP} {\bf 0502} (2005) 002,
  [\href{http://xxx.lanl.gov/abs/hep-ph/0411254}{{\tt hep-ph/0411254}}].

\bibitem{Birkedal-Hansen:2003mp}
A.~Birkedal-Hansen and J.~G. Wacker, {\em Phys. Rev.} {\bf D69} (2004) 065022,
  [\href{http://xxx.lanl.gov/abs/hep-ph/0306161}{{\tt hep-ph/0306161}}].

\bibitem{Hubisz:2004ft}
J.~Hubisz and P.~Meade, {\em Phys. Rev.} {\bf D71} (2005) 035016,
  [\href{http://xxx.lanl.gov/abs/hep-ph/0411264}{{\tt hep-ph/0411264}}].

\bibitem{Martin:2006ss}
A.~Martin, \href{http://xxx.lanl.gov/abs/hep-ph/0602206}{{\tt hep-ph/0602206}}.

\bibitem{Belanger:2001fz}
G.~Belanger, F.~Boudjema, A.~Pukhov, and A.~Semenov, {\em Comput. Phys.
  Commun.} {\bf 149} (2002) 103--120,
  [\href{http://xxx.lanl.gov/abs/hep-ph/0112278}{{\tt hep-ph/0112278}}].

\bibitem{Belanger:2004yn}
G.~Belanger, F.~Boudjema, A.~Pukhov, and A.~Semenov, {\em Comput. Phys.
  Commun.} {\bf 174} (2006) 577--604,
  [\href{http://xxx.lanl.gov/abs/hep-ph/0405253}{{\tt hep-ph/0405253}}].

\bibitem{Gondolo:2004sc}
P.~Gondolo {\em et.~al.}, {\em JCAP} {\bf 0407} (2004) 008,
  [\href{http://xxx.lanl.gov/abs/astro-ph/0406204}{{\tt astro-ph/0406204}}].

\bibitem{Baer:2002fv}
H.~Baer, C.~Balazs, and A.~Belyaev, {\em JHEP} {\bf 03} (2002) 042,
  [\href{http://xxx.lanl.gov/abs/hep-ph/0202076}{{\tt hep-ph/0202076}}].

\bibitem{Pukhov:2004ca}
A.~Pukhov, \href{http://xxx.lanl.gov/abs/hep-ph/0412191}{{\tt hep-ph/0412191}}.

\bibitem{Semenov:2002jw}
A.~V. Semenov, \href{http://xxx.lanl.gov/abs/hep-ph/0208011}{{\tt
  hep-ph/0208011}}.

\bibitem{Birkedal:2006fz}
A.~Birkedal, A.~Noble, M.~Perelstein, and A.~Spray,
  \href{http://xxx.lanl.gov/abs/hep-ph/0603077}{{\tt hep-ph/0603077}}.

\bibitem{Brun:2006cj}
P.~Brun, \href{http://xxx.lanl.gov/abs/astro-ph/0603387}{{\tt
  astro-ph/0603387}}.

\bibitem{Skands:2003cj}
P.~Skands {\em et.~al.}, {\em JHEP} {\bf 07} (2004) 036,
  [\href{http://xxx.lanl.gov/abs/hep-ph/0311123}{{\tt hep-ph/0311123}}].

\bibitem{Gelmini:1990je}
G.~B. Gelmini, P.~Gondolo, and E.~Roulet, {\em Nucl. Phys.} {\bf B351} (1991)
  623--644.

\bibitem{Edsjo:1997bg}
J.~Edsjo and P.~Gondolo, {\em Phys. Rev.} {\bf D56} (1997) 1879--1894,
  [\href{http://xxx.lanl.gov/abs/hep-ph/9704361}{{\tt hep-ph/9704361}}].

\bibitem{Feng:2003uy}
J.~L. Feng, A.~Rajaraman, and F.~Takayama, {\em Phys. Rev.} {\bf D68} (2003)
  063504, [\href{http://xxx.lanl.gov/abs/hep-ph/0306024}{{\tt
  hep-ph/0306024}}].

\bibitem{Belanger:prep}
G.~Belanger, F.~Boudjema, P.~Brun, A.~Pukhov, S.~Rosier-Lees, P.~Salati, and
  A.~Semenov, \href{http://xxx.lanl.gov/abs/in preparation}{{\tt in
  preparation}}.

\bibitem{Eidelman:2004wy}
S.~Eidelman {\em et.~al.},, {\bf Particle Data Group} Collaboration {\em Phys.
  Lett.} {\bf B592} (2004) 1.

\bibitem{Djouadi:2002ze}
A.~Djouadi, J.-L. Kneur, and G.~Moultaka,
  \href{http://xxx.lanl.gov/abs/hep-ph/0211331}{{\tt hep-ph/0211331}}.

\bibitem{Allanach:2001kg}
B.~C. Allanach, {\em Comput. Phys. Commun.} {\bf 143} (2002) 305--331,
  [\href{http://xxx.lanl.gov/abs/hep-ph/0104145}{{\tt hep-ph/0104145}}].

\bibitem{Porod:2003um}
W.~Porod, {\em Comput. Phys. Commun.} {\bf 153} (2003) 275--315,
  [\href{http://xxx.lanl.gov/abs/hep-ph/0301101}{{\tt hep-ph/0301101}}].

\bibitem{Paige:2003mg}
F.~E. Paige, S.~D. Protopescu, H.~Baer, and X.~Tata,
  \href{http://xxx.lanl.gov/abs/hep-ph/0312045}{{\tt hep-ph/0312045}}.

\bibitem{Lee:2003nt}
J.~S. Lee {\em et.~al.}, {\em Comput. Phys. Commun.} {\bf 156} (2004) 283--317,
  [\href{http://xxx.lanl.gov/abs/hep-ph/0307377}{{\tt hep-ph/0307377}}].

\bibitem{Choi:2004rf}
S.~Y. Choi, M.~Drees, and B.~Gaissmaier, {\em Phys. Rev.} {\bf D70} (2004)
  014010, [\href{http://xxx.lanl.gov/abs/hep-ph/0403054}{{\tt
  hep-ph/0403054}}].

\bibitem{Chang:1998uc}
D.~Chang, W.-Y. Keung, and A.~Pilaftsis, {\em Phys. Rev. Lett.} {\bf 82} (1999)
  900--903, [\href{http://xxx.lanl.gov/abs/hep-ph/9811202}{{\tt
  hep-ph/9811202}}].

\bibitem{Pilaftsis:2002fe}
A.~Pilaftsis, {\em Nucl. Phys.} {\bf B644} (2002) 263--289,
  [\href{http://xxx.lanl.gov/abs/hep-ph/0207277}{{\tt hep-ph/0207277}}].

\bibitem{Hagiwara:2002fs}
K.~Hagiwara {\em et.~al.},, {\bf Particle Data Group} Collaboration {\em Phys.
  Rev.} {\bf D66} (2002) 010001.

\bibitem{Allanach:2005kk}
B.~C. Allanach {\em et.~al.},. Presented at Les Houches Workshop on Physics at
  TeV Colliders, Les Houches, France, 2-20 May 2005.

\bibitem{Ellwanger:2005dv}
U.~Ellwanger and C.~Hugonie, \href{http://xxx.lanl.gov/abs/hep-ph/0508022}{{\tt
  hep-ph/0508022}}.

\bibitem{Ellwanger:2004xm}
U.~Ellwanger, J.~F. Gunion, and C.~Hugonie, {\em JHEP} {\bf 02} (2005) 066,
  [\href{http://xxx.lanl.gov/abs/hep-ph/0406215}{{\tt hep-ph/0406215}}].

\bibitem{Pukhov:1999gg}
A.~Pukhov {\em et.~al.}, \href{http://xxx.lanl.gov/abs/hep-ph/9908288}{{\tt
  hep-ph/9908288}}.

\end{thebibliography}

\providecommand{\href}[2]{#2}\begingroup\raggedright\endgroup

\end{document}